\documentclass[preprint, double-spaced]{revtex4}
\usepackage{graphicx}
\usepackage{amsmath,bbm}
\usepackage{amssymb,bm}

\begin{document}

\title{Prediction of Charged Hadron Multiplicity at LHC and CBM Experiments}
\author{}
\author{Ashwini~Kumar}
\author{B.~K.~Singh}
\author{P.~K.~Srivastava\footnote{corresponding author: $prasu111@gmail.com$}}
\author{C.~P.~Singh}

\affiliation{Department of Physics, Banaras Hindu University, 
Varanasi 221005, INDIA}

\begin{abstract}
\noindent
A systematic study of charged hadron multiplicities ($n_{ch}$) at various collision energies is very much important in understanding the basic production mechanism of the hadrons in nucleus-nucleus collision experiments. Furthermore, the variations of $n_{ch}$ in nucleus-nucleus collisions with respect to the colliding energy and mass number can provide a potential probe for the formation of quark gluon plasma (QGP) in the laboratory. In this paper, we propose a phenomenological model based on the constituent quark-quark interactions to calculate the average multiplicity ($n_{ch}$) and pseudorapidity density at mid-rapidity ($(dn_{ch}/d\eta)_{\eta=0}$) of charged hadrons at various center-of-mass energies $(\sqrt{s_{NN}})$ for nucleus-nucleus ($A-A$) collisions. We first propose a new parametrization for $n_{ch}^{pp}$ and $(dn_{ch}/d\eta)^{pp}_{\eta=0}$ in $p-p$ interactions based on some initial inputs which fit the experimental data very well. We further extend this parametrization by using simple phenomenological assumptions regarding mean number of participating quarks and mean number of collisions to obtain the $n_{ch}$ and $(dn_{ch}/d\eta)_{\eta=0}$ for $A-A$ collisions and show their dependencies on the mass number of colliding nuclei as well as on  $\sqrt{s_{NN}}$. We also compare the results obtained from our model with the results obtained from the modified Glauber model in order to demonstrate the difference between the two formalisms. Finally, we compare the charged hadron multiplicity and pseudorapidity density at mid-rapidity for $A-A$ collisions obtained from our model with the available experimetal data from various heavy-ion collision experiments and give our predictions for $A-A$ collisions at the Large Hadron Collider (LHC) and at Compressed Baryonic Matter (CBM) experiments.   
\\


\end{abstract}
\maketitle 
\section{Introduction}
The ultimate goal of ultra-relativistic heavy-ion collision experiments is to test the predictions of quantum chromodynamics (QCD), which is unanimously believed to be the theory of strong interactions [1]. The ultra-relativistic nuclear collisions reveal the nature of hadronic interactions at very short time and/or distance and throw light on the role played by the internal structure of hadrons in the multiparticle production. Intensive theoretical efforts have also continued for more than three decades to understand the particle production mechanism in the nucleus-nucleus interactions. Hadronic multiplicities and their correlations can reveal information on the nature, composition, and size of the fireball from which they are originating [2]. Of particular interest is the limit up to which the chemical/thermal equilibration is possible in the fireball. The appearance of quark gluon plasma (QGP) which involves a partonic medium at local thermodynamic equilibrium, and its subsequent hadronization after the phase transition, should in general drive hadrons also towards chemical equilibrium. The theoretical description of particle production depends essentially on two ingredients :  many-body interactions and the understanding of the production processes. 

 The search for some regularities and systematics in the multiplicity distributions in hadron-hadron, hadron-nucleus and nucleus-nucleus collisions is a fascinating topic because it hints at the underlying production mechanism. However, our understanding of the subject is still very poor partially because of the reason that in soft hadronic or nuclear collisions, the role of perturbative QCD is less clear. Several new experimental information on multiparticle production have accumulated in recent years. Consequently several theoretical approaches have been developed in an attempt to understand or organise the data. Several efforts [3-8], which are based on hydrodynamic descriptions, have also been proposed to explain the enormous experimental data coming from a variety of collision experiments ranging from very low energy (e.g., SIS) to very high energy (e.g., LHC). In addition to this, several approaches based on statistical production of particles have recently appeared in the literature [9-23]. The statistical models are based on the assumption of local filling of available phase space according to statistical laws. Furthermore, there are models which try to explain the rather complicated heavy ion colliding processes by analyzing them in terms of the basic quark-gluon interaction processes e.g., additive quark model (AQM) [24], dual parton model (DPM) [25] or colour neutralization model (CNM) [26] etc. The main difference in these models lies in the assumption about the potential number of participant partons which can interact independently. Traditional AQM treats the elastic scattering of two hadrons at high energy in terms of Pomeron exchange between two quarks involving one from each hadron i.e., the number of participants is limited to the valance quarks only [24, 27]. However, in DPM and CNM, the number of participant quarks is unlimited because these can involve the sea quarks as well. Many attempts have also been made in the past for searching certain systematics or scaling relations which are universal to all types of reactions, i.e., lepton-hadron, hadron-hadron, hadron-nucleus and nucleus-nucleus collisions. Any distinct deviation from these relations observed in the ultrarelativistic nuclear collisions will be a potential indicator of a new and exotic phenomenon occuring there [1]. In order to find out whether deviations arise from the presence of some exotic phenomena, e.g., formation of QGP, where we must know the background arising from nucleus-nucleus collisions without the presence of any such phase transition [1]. In this paper, we propose a model for the multiparticle production and search hints for deviations signaling an exotic phenomenon by comparing our results with the experimental data coming from a variety of heavy-ion collision experiments. Our model involves a main assumption that the nucleus-nucleus collision can be considered as a superposition of independent quark-quark collisions, the number of which is determind by the geometry of the collisions.  

Bascially in QCD, the interaction mechanism between target and projectile nuclei can be described as follows : A projectile quark exchanges a gluon with a target quark and colour forces are thus stretched between them as well as other constituents because they try to restore the colour singlet behaviour. When two quarks thus separate, the colour force builds up a field between them and as the energy in the colour field increases, the colour tubes break up into hadrons and quark-antiquark pairs are created. In this paper, we have basically used the model given by Singh et. al. [28] and obtain the charged hadron multiplicities for $p-p$, $h-A$, and $A-A$ collisions. This model considers a multiple collision scheme in which a valance quark of the incident nucleon suffers one or more inelastic collisions with a valance quark of the target nucleon. The quark thus loses energy and momenta and produces hadrons in each quark-quark collision. However, all the collisions are independent and their effects should be incoherently superimposed. Further in $A-A$ collisions, one has to adequately incorporate the number of participating quarks and mean number of quark-quark collisions. The rest of the paper is organized as follows : In section II, we will give the detailed description of the model used in this paper. In section III, we give the results and the discussion and finally in section IV, we will give the conclusions based on this work and the future prospects of this study.

\section{Description of Model}
If we assume a universal mechanism of charged particle production in the hadron-hadron, hadron-nucleus and nucleus-nucleus collisions, it must be driven by the available amount of energy involved for the secondary production, and it must depend on the mean number of participant quarks. The main ingredients of our model is based on a phenomenological model proposed earlier by Singh et. al. [28]. Let us attempt first to understand the basic particle production mechanism in $p-p$ collisions. There are several papers emphasizing various types of fitting parametrizations in order to provide a unified description for the produced charged particles in $p-p$ collisions at various energies. We have taken some of these functions with the values of their constants from the literature and fitted them with the experimental data. All these parametrizations fit the experimental data for $p-p$ collisions at intermediate and higher energies but usually show some disagreement with the experimantal data at lower energies ( see Fig. 1). Therefore, we propose here a new parametrization to accomodate entire $p-p$ experimental data [29-38] from low energies to very high energies (i.e., from 6.15 GeV upto 7 TeV), in a unified way as follows :
\begin{equation}
 <n_{ch}> _{pp}=(a'+b' ln \sqrt{s_{a}}+c'ln^{2} \sqrt{s_{a}}+d'ln^{3} \sqrt{s_{a}})-\alpha.
\end{equation}
In Eq. (1), $\alpha$ is the leading particle effect and $\sqrt{s_{a}}$ is the available center-of-mass energy i.e., $\sqrt{s_{a}}=\sqrt{s}-m_{B}-m_{T}$, where $m_{B}$ is the mass of projectile and $m_{T}$ the mass of the target nucleon, respectively; $a'$, $b'$, $c'$ and $d'$ are constants. We find the values $a'=1.8$, $b'=0.37$, $c'=0.43$ and $d'=0.04$ as derived from the best fit to the data [29-38]. 

Now, we can extrapolate this expression for the produced charged particles in the hadron-nucleus interaction based on two basic assumptions. First assumption is that the number of constituent quarks which participated in hadron-nucleus collisions share the total available center-of-mass energy $\sqrt{s_{A}}$ and thus the energy available to each interacting quark becomes $\sqrt{s_{A}}/N_{q}$ in $h-A$ collisions. Now, since $s_{a}$ is the available center-of-mass energy in $h-p$ collision for one effective collision of quarks of hadronic beam in the target, therefore, second assumption is that the total available square of center-of-mass energy $s_{A}$ in $h-A$ case becomes $\nu_{q}s_{a}$ provided each quark suffers on average $\nu_{q}$ collisions. Thus, the expression for average charged hadron multiplicity in $h-A$ collisions can be expressed as follows [28]:
\begin{equation}
 <n_{ch}> _{hA}=N_{q}\left[a'+b' ln \left(\frac{\sqrt{s_{A}}}{N_{q}}\right)+c'ln^{2}\left(\frac{\sqrt{s_{A}}}{N_{q}}\right) +d'ln^{3}\left(\frac{\sqrt{s_{A}}}{N_{q}}\right) \right]-\alpha.
\end{equation}
In the above Eq. (2), $\sqrt{s_{A}}= (\nu_{q}s_{a})^{1/2}$, where $\nu_{q}$ is the mean number of inelastic quark collisions in the target nucleus and is defined as : $\nu_{q}=A\sigma_{qN}^{in}/\sigma_{qA}^{in}$. Here $A$ is the mass number of the target nucleus and $\sigma_{qN}^{in}$ is the quark-nucleon inelastic interaction cross-section and equal to one-third of nucleon-nucleon inelastic cross-section ($\sigma_{NN}^{in}$) i.e., $\sigma_{qN}^{in}\approx (1/3)\sigma_{NN}^{in}$ [17]. Also $\sigma_{qA}^{in}$ ,the quark-nucleus inelastic interaction cross-section and is obtained from Glauber's approximation by using the following expression [18]:
\begin{equation}
\sigma_{qA}^{in}=\int d^{2}b\left[1-\left(1-\sigma_{qN}^{in}D_{A}(b)\right)^{A}\right],
\end{equation}
where profile function $D_{A}(b)$ is related to nuclear density, $\rho(b,z)$ by the relation :
\begin{equation}
D_{A}(b)=\int_{-\infty}^{\infty}\rho(b,z)dz.
\end{equation}
We use the following expression to calculate the nuclear density :
\begin{equation}
\rho(b,z)=\frac{\rho_{0}}{1-exp(\frac{\sqrt{b^{2}+z^{2}}-R}{a})},
\end{equation}
where $R$ and $a$ are constants for any nucleus having mass number $A$, $\rho_{0}$ is the normalization constant and $b$ is the impact parameter. We have taken the values of the constants from the Ref. [39]. Furthermore, $N_{q}$ in Eq. (3), is the mean number of participant quarks and is defined as follows :
\begin{equation}
N_{q}=\frac{N_{c}\sigma_{qA}^{in}}{\sigma_{hA}^{in}},
\end{equation}
where $N_{c}$ is the number of valance quarks in the nucleus A. 
\\
The generalization of the above picture for the case of nucleus-nucleus collisions goes along the same line and can be given as follows :
\begin{equation}
<n_{ch}> _{AB}=N_{q}^{AB}\left[a'+b' ln \left(\frac{\sqrt{s_{AB}}}{N_{q}^{AB}}\right)+c'ln^{2}\left(\frac{\sqrt{s_{AB}}}{N_{q}^{AB}}\right) +d'ln^{3}\left(\frac{\sqrt{s_{AB}}}{N_{q}^{AB}}\right) -\alpha\right],
\end{equation}
where $\sqrt{s_{AB}}=A(\nu_{q}^{AB}s_{a})^{1/2}$ and the mean number of inelastic quark collision $\nu_{q}^{AB}$ can be given as follows :
\begin{equation}
\nu_{q}^{AB}=\nu_{qA}\nu_{qB}=\frac{A\sigma_{qN}^{in}}{{\sigma_{qA}^{in}}}.\frac{B\sigma_{qN}^{in}}{{\sigma_{qB}^{in}}}.
\end{equation}
Furthermore, mean number of participating quarks $N^{AB}_{q}$ can be calculated by generalizing Eq. (6) in the following manner:
\begin{equation}
N^{AB}_{q}=\frac{1}{2}\left[\frac{N_{B}\sigma_{qA}^{in}}{{\sigma_{AB}^{in}}}+\frac{N_{A}\sigma_{qB}^{in}}{{\sigma_{AB}^{in}}}\right],
\end{equation}
where $\sigma_{AB}^{in}$ is the inelastic cross-section for nucleus A- nucleus B collision and can be expressed in the following manner [18]:
\begin{equation}
\sigma_{AB}^{in}= \pi r^{2}\left[A^{1/3}+B^{1/3}-\frac{c}{A^{1/3}+B^{1/3}}\right]^2,
\end{equation}
where c is a constant and has a value 4.45 for nucleus-nucleus collisions. 
One can see that the parametrization as given by Eq. (7) gives the most general relation relating nucleus-nucleus collisions to hadron-nucleus and hadron-proton collisions and the values of the parameters $a', b', c'$, and $d'$ remain unaltered which shows the similarity in the role of basic quark-gluon interaction in all these processess.
In a search for creating quark gluon plasma (QGP), greater emphasis is laid on the central or head-on collisions of two nuclei. In such a case, we assume that all the quarks of the beam nucleus are wounded and the resulting mean multiplicity can be obtained by using Eq. (7) in the following manner :
\begin{equation}
<n_{ch}>^{central}_{AB}=A\left[a'+b'ln(\nu_{q}^{AB}s_{a})^{1/2}+c'ln^{2}(\nu_{q}^{AB}s_{a})^{1/2}+d'ln^{3}(\nu_{q}^{AB}s_{a})^{1/2}-\alpha\right].
\end{equation}

 The pseudorapidity distribution of charged particles is another very important quantity in the studies of particle production mechanism in high energy $h-h$ and $A-A$ collisions. It has been pointed out that $(dn_{ch}/d\eta)$ can be used to obtain the temperature ($T$) and density ($\rho$) of the QGP [40-42]. Moreover, mid-rapidity density reflects the different stages of the reaction. To calculate the pseudorapidity density of charged hadrons, we first fit the experimental data of $(dn_{ch}/d\eta)^{pp}_{\eta=0}$ by using similar parametrization as was used in Eq. (1) but with slightly different values of constants. Further extrapolation of two component model [43-45] gives $(dn_{ch}/d\eta)^{AA}_{\eta=0}$ as follows :
\begin{equation}
\left(\frac{dn_{ch}}{d\eta}\right)^{AA}_{\eta=0}=\left(\frac{dn_{ch}}{d\eta}\right)^{pp}_{\eta=0}\left[\left(1-x\right)N_{q}^{AB}+ x N_{q}^{AB}\nu_{q}^{AB}\right],
\end{equation}
where $x$ quantifies the relative contribution from hard and soft processes. The fraction $x$ corresponds to the hard processes and the remaining fraction ($1-x$) arising from soft processes. Thus the contribution from hard processes is propotional to the total number of collisions i.e., $N_{q}^{AB}\nu_{q}^{AB}$ and the portion from soft processes is propotional to the participating quark constituents i.e., $N_{q}^{AB}$.

\begin{figure}
 \includegraphics[height=30em]{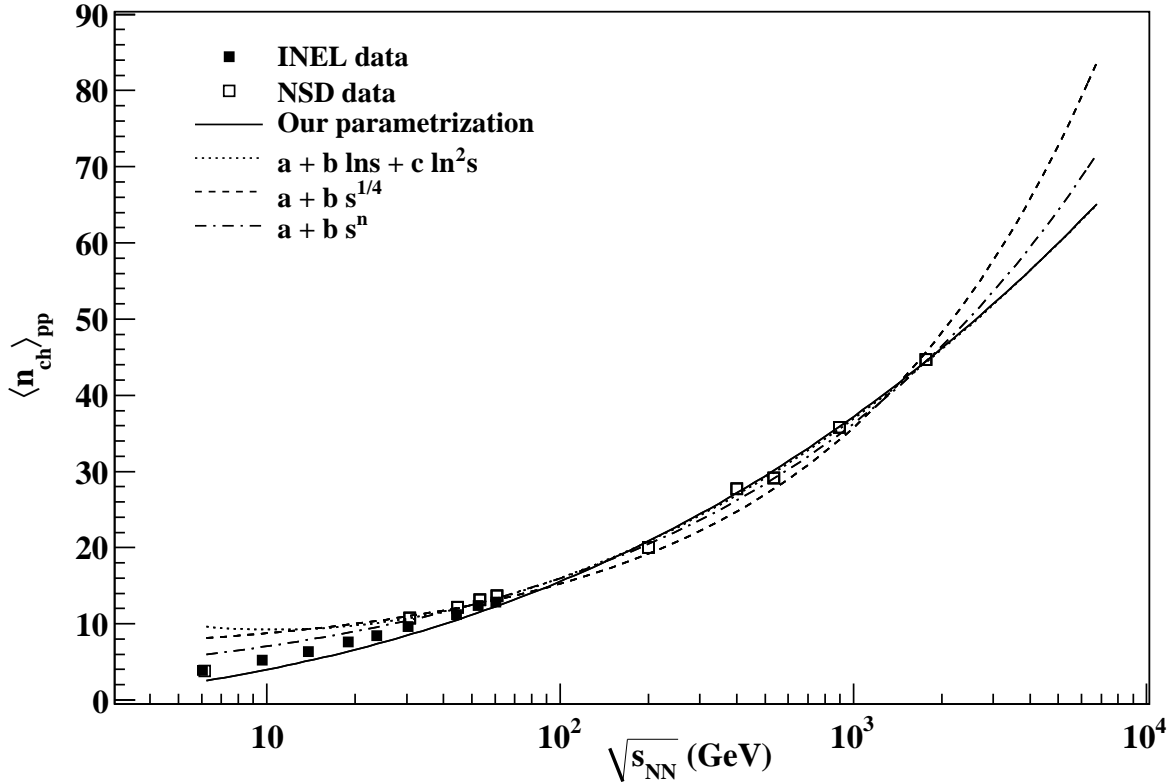}%use pdflatex for pdf
\caption[]{Variation of total multiplicities of charged hadrons in $p-p$ collision at various $\sqrt{s_{NN}}$.}
\end{figure}

\section{Results and Discussions}
 \begin{figure}
 \includegraphics[height=30em]{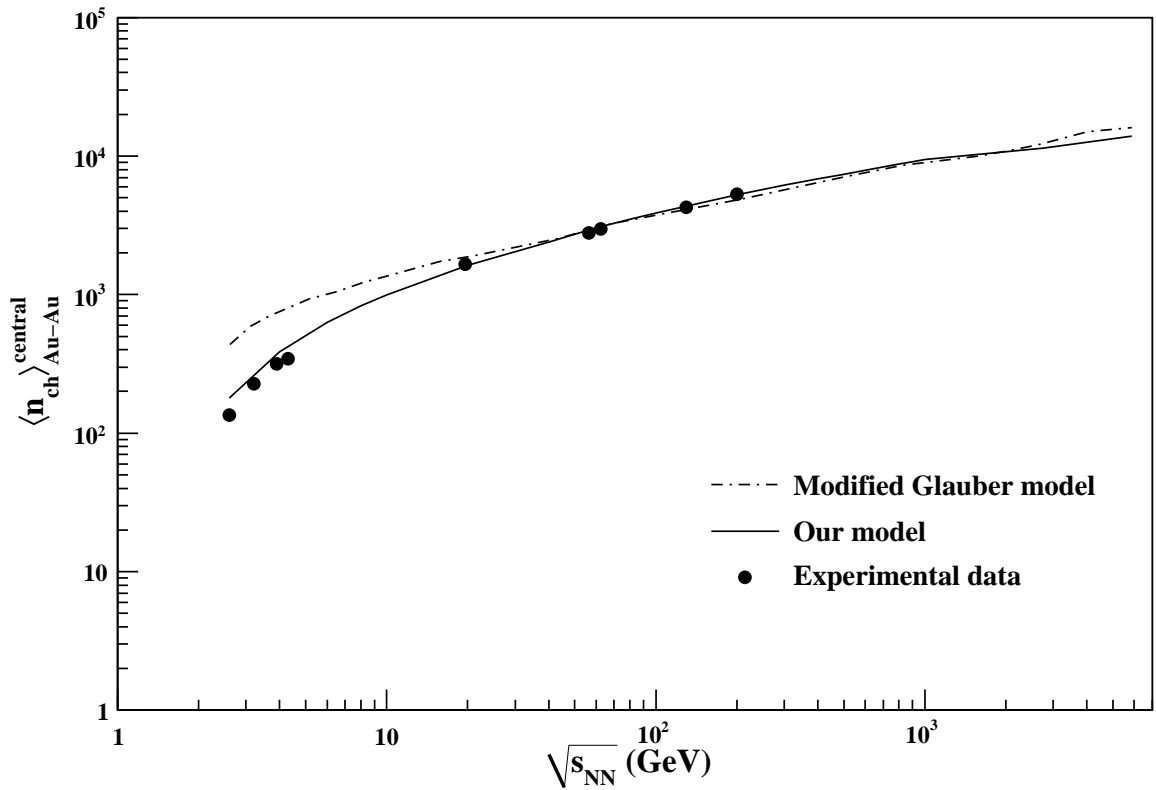}%use pdflatex for pdf
\caption[]{Variation of total mean multiplicity of charged hadrons in central $Au-Au$ collisions at various $\sqrt{s_{NN}}$.}
\end{figure}

In Fig. (1), we present the inelastic (filled symbols) and non-single diffractive (NSD) data (open symbols) of charged hadron multiplicity in full phase space for $p-p$ collisions at various center-of-mass energies from different experiments e.g., ISR, UA5 and E735 [29-38]. We use inelastic data at very low energies (filled symbols) because there is no NSD data available for these energies and also the trend shows that the difference between inelastc and NSD data will be very small at lower energies. Further, we fit this data set with four different functional forms. The short-dashed line has the functional form as : $a+bs^{1/4}$ which is actually inspired by the Fermi-Landau model [46, 47]. It provides a resonable fit to the data at higher $\sqrt{s_{NN}}$ with $a = 5.774$ and $b=0.948$ [29]. However, since $a$ summarizes the leading particle effect, it should not be much larger than two. The dotted line has the functional form as : $a+b$ $ln s+c$ $ln^{2}s$ and it fits the data well at higher $\sqrt{s_{NN}}$ but shows a disagreement with the experimental data at lower center-of-mass energies. The dashed-dotted line represents the form $a+bs^{n}$ and it also provides a good qualitative description of the data with $a=0, ~b=3.102$ and $n=0.178$ [29]. The solid line represents our parametrization given by Eq. (1) and it undoubtedly represents the most reasonable good fit to the data starting from very low upto very high $\sqrt{s_{NN}}$.

\begin{figure}
 \includegraphics[height=30em]{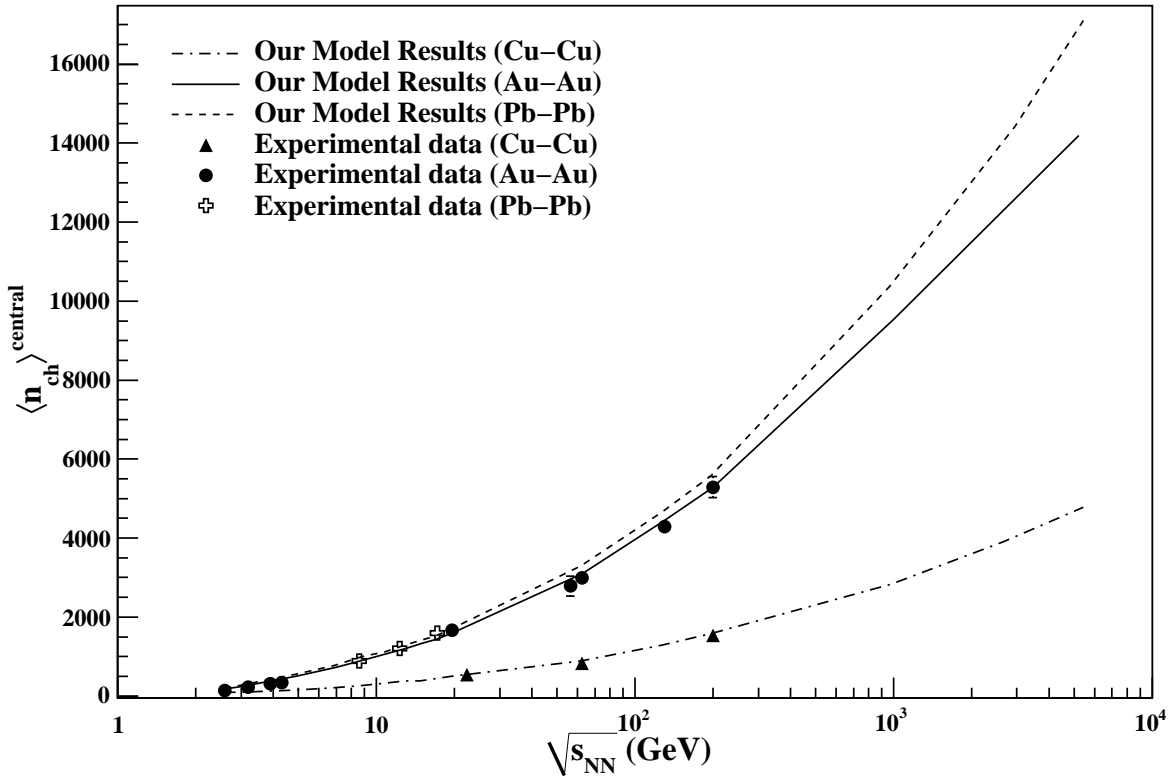}%use pdflatex for pdf
\caption[]{Variation of total mean multiplicity of charged hadrons in central collisions for different colliding nuclei at various $\sqrt{s_{NN}}$.}
\end{figure}

Fig. 2, shows the variation of mean multiplicity of charged hadrons produced in central $Au-Au$ collision with respect to $\sqrt{s_{NN}}$. We also compare our model results (solid line) and the modified Glauber model results (dash-dotted line) [48] with the experimental data of AGS and RHIC [49-52]. We would like to mention here that the main difference in modified Glauber model with respect to the standard Glauber calculations is that for each nucleon-nucleon collision modified model uses the value of $\sigma^{pp}_{inel}$ at the corresponding centre-of-mass nucleon-nucleon collision energy. We find that the results obtained from our model give a good description to the experimental data in comparison to the modified Glauber model predictions, especially at lower center-of-mass energies.

\begin{table}

\begin{center}
\caption{Total mean multiplicity of charged hadrons produced in $d-Au$ collision at $\sqrt{s_{NN}}=200$ GeV.}
\resizebox{9cm}{!}{
\begin{tabular}{cccc} \hline \hline 
                     &Colliding Nuclei     & $\langle n_{ch}\rangle^{min. bias}$        & $\langle n_{ch}\rangle^{central}$                \\ 
         
\hline 
Our Model            & $d-Au$               & $82$                             & $162$      \\
PHOBOS [54]         & $d-Au$               & $87_{-6}^{+7}$                 & $167_{-11}^{+14}$    \\ \hline
\end{tabular}
}
\end{center}
%\hspace{1cm}
\end{table} 

In Fig. 3, we show the variation of mean-multiplicity calculated from our model for $Au-Au$, $Pb-Pb$ and $Cu-Cu$ central collisions with respect to $\sqrt{s_{NN}}$. We also compare our model results with the experimantal data of RHIC and SPS experiments [49-53]. One can see from Fig. 3, that the model results give excellent fit to the separate experimental data points for $Cu-Cu$,  $Au-Au$ and $Pb-Pb$, respectively. This concludes that our phenomenological model works well because it considers the picture of incoherent superposition of basic quark-quark collisions in describing the nucleus-nucleus collisions.

In Table I, we have shown the mean multiplicity of charged hadrons in minimum-bias as well as in central $d-Au$ collision at $\sqrt{s_{NN}}=200$ GeV. We have also shown the experimental results [54] for this collision for comparison. The good agreement between the experimental and our results shows that our model can be used for asymmetric colliding nuclei as well. 

\begin{figure}
 \includegraphics[height=30em]{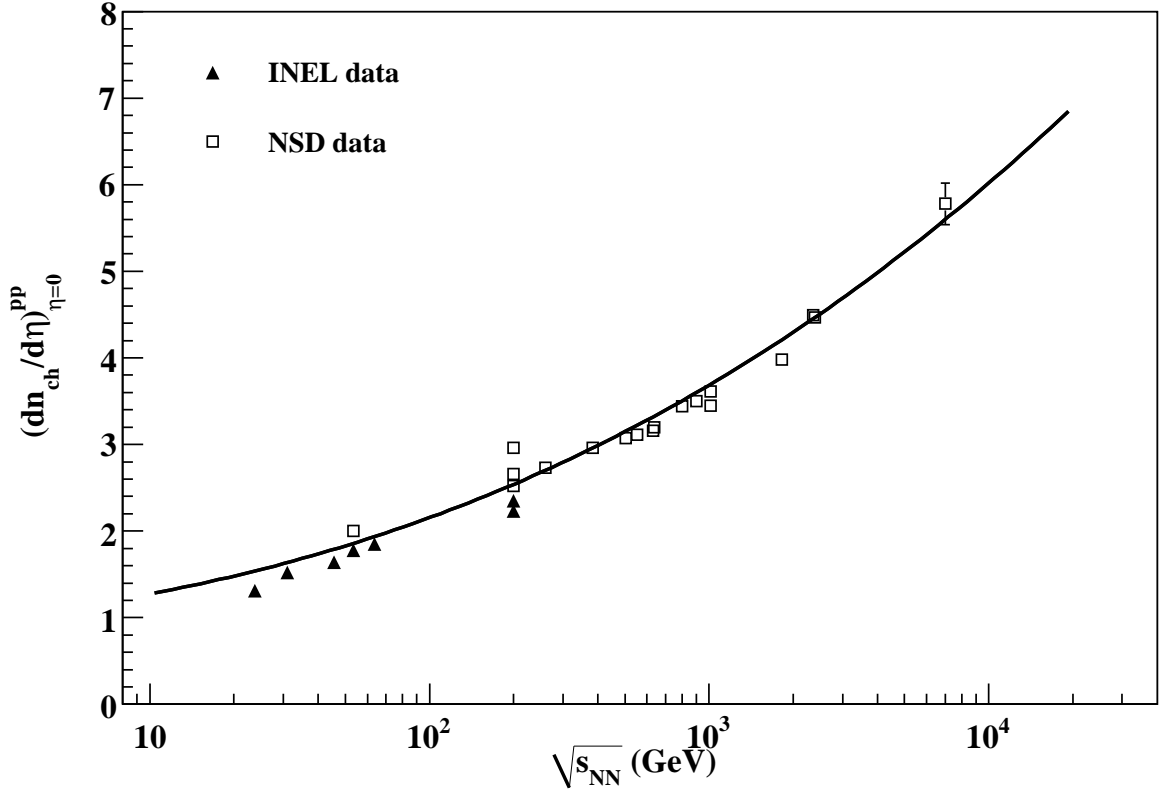}%use pdflatex for pdf
\caption[]{Variation of pseudorapidity density at mid-rapidity for $p-p$ collisionat various $\sqrt{s_{NN}}$. Filled symbols are data from inelastic $p-p$ events [32, 53-54] and open symbols are experimental data for NSD events [33-36, 58-59]. Solid line is the outcome of our parametrization.}
\end{figure}

In Fig. 4, we present the inelastic (filled symbols) and non-single diffractive (NSD) data (open symbols) of $(dn_{ch}/d\eta)_{\eta=0}$ for $p-p$ collisions at various center of mass energies from different experiments e.g., ISR, UA5, E735, RHIC and LHC [32, 33-36, 55-59]. Further, we fit (solid line) this data set with  our functional form of Eq. (1) and obtained the values of constants as $a' = 1.78$, $b' =0.025$, $c' =0.041$ and $d' =0.0017$. The central rapidity hadron density thus continues to increase with the colliding energy at LHC in a triple logarithmic form.

\begin{figure}
 \includegraphics[height=30em]{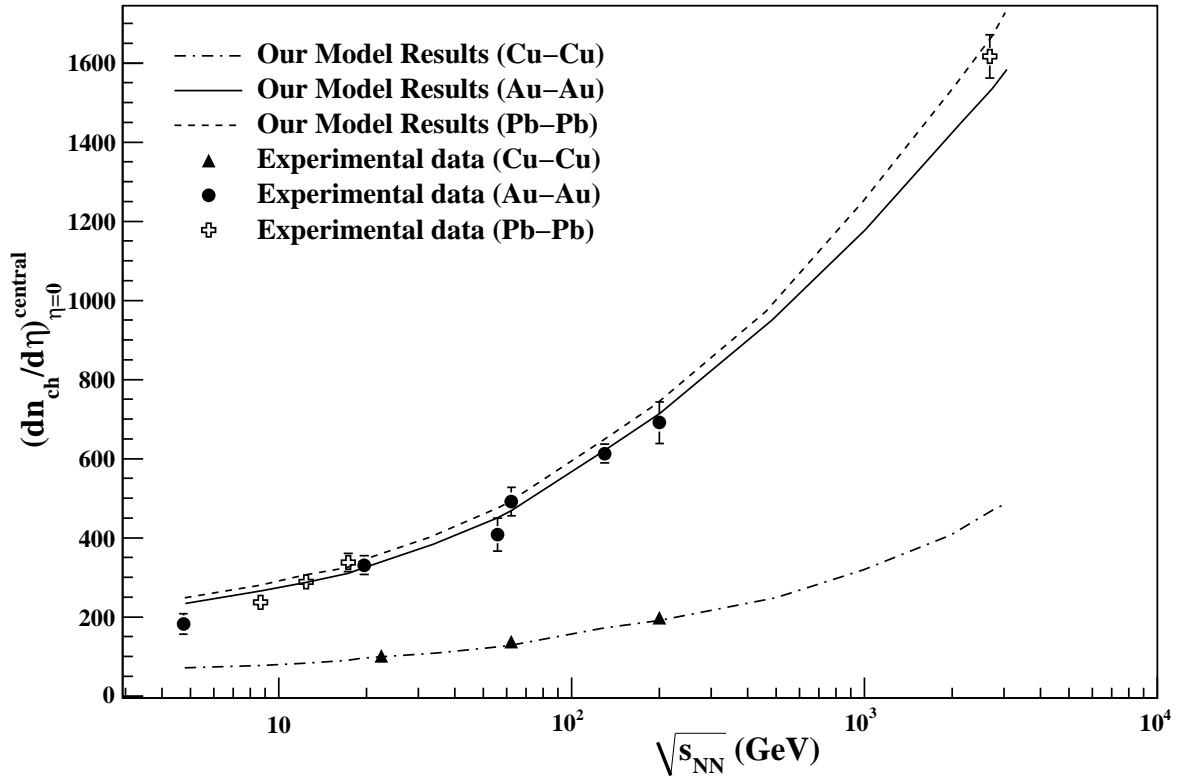}%use pdflatex for pdf
\caption[]{Variation of pseudorapidity density at mid-rapidity for $A-A$ collisionat various $\sqrt{s_{NN}}$. Symbols are data from various heavy ion collision experiments [49-53, 60-61].}
\end{figure}

In Fig. 5, we plot the variation of $(dn_{ch}/d\eta)_{\eta=0}$ obtained in our model for $Au-Au$, $Pb-Pb$ and $Cu-Cu$ central collisions with respect to $\sqrt{s_{NN}}$. We also compare our model results with the experimental data of RHIC, SPS and LHC experiments [49-53, 60-61]. Our model results compare well with the experimental data from very low energy experiments to very high energy experiments. As we didn't take any effect of final-state interactions in our model, it verifies that the hadron multiplicity in $A-A$ collisions is mainly driven by initial parton production and the effect of final-state interaction is negligibly small as suggested in some earlier models [62-63].

\begin{table}

\begin{center}
\caption{Predictions of total mean multiplicity of charged hadrons produced in heavy ion collision experiments.}
\resizebox{16.5cm}{!}{
\begin{tabular}{ccccccc} \hline \hline 
Experiment             &Colliding Nuclei   & $\sqrt{s_{NN}}$   & $\langle n_{ch}\rangle^{min. bias}_{our ~model}$     & $\langle n_{ch}\rangle^{central}_{our ~model}$    &  $\langle n_{ch}\rangle^{central}_{Ref. [66-67]}$       &  $\langle n_{\pi}\rangle^{central}_{HSD [64-65]}$     \\ 
             &                & $(GeV)$ &  &  &   & \\ 
\hline
LHC          & $Pb-Pb$          & $2760$                             & $8895$    & $14377$    & $15000\pm 1000$   &  --\\
CBM          & $Au-Au$          & $3.97$ ($E_{lab}\approx 8 A$)      & $328.75$  & $380.8$    & --                & $308$\\
 ''          & $Au-Au$          & 8.1  ($E_{lab}\approx 35 A$)     & $670.8$   & $837$      & --       & $650$ \\ \hline
\end{tabular}
}
\end{center}
%\hspace{1cm}
\end{table}

In Table.II, we give the predictions for the mean multiplicity of charged hadrons, produced in minimum-bias events as well as in central events, based on our model calculations at LHC and CBM energies. Specifically, we calculate the charged hadron multiplicity in $Pb-Pb$ collisions at $\sqrt{s_{NN}}= 2.76$ TeV for LHC and the mean multiplicity of charged hadrons in $Au-Au$ collisions at $E_{lab}= 8$ AGeV and $35$ AGeV for future CBM experiment. We also compare our model results with the other model predictions like hadron string dynamics (HSD) model predictions for CBM experiment [64-65] and model prediction of W. Busza for LHC experiment [66-67]. 

\section{Conclusions and Future Prospects}
We have given a parametrization which correctly describes the multiplicity distributions in particle and nuclear collisions from a few GeV upto the highest LHC energies. It involves the interactions of the constituent quarks of the colliding beams or beam and target. The collision of the two Lorentz-contracted particles results in a complete thermalized system. We assume that the production of secondary particles is proportional to the fraction of the available energy for the participant quarks. This picture describes consistently the `soft' hadron production in $p-p$, $p-\bar{p}$, $A-A$ and $p-A$ collisions. In heavy-ion collisions, more than one quark per nucleon interacts due to the large size of the nucleus as well as due to a large travel path to be travelled inside the nucleus [68]. For more central collisions, larger number of interactions occur and hence we get larger energy available for the secondary particle production. In the most central nucleus-nucleus collisions, all three constituent quarks from each nucleon will interact simultaneously and they deposit their energy into the thermalized collisions volume. We find that our parametrization gives the excellent fit to the $p-p$ data for the entire energy range. We also find that a suitable extension of this description works very well for the $A-A$ collisions again for the entire energy range. We have also compared our model results with the results obtained from modified Glauber model to demonstrate the difference between nucleon-nucleon interaction and quark-quark interaction pictures. We have shown the charged particle multiplicity in central as well as in minimum-bias $d-Au$ collisions at $\sqrt{s_{NN}}=200$ GeV and we find it compares well with the experimental result. This exercise shows that our model works well for assymetric colliding nuclei case also. We have also proposed an extension of the two-component model based on our model and attempted to get the rapidity distribution and we find that it again works well. We have given our prediction for LHC and CBM data as well and compare them with some recent model predictions [64-67]. 

 In conclusion, we have attempted to draw certain universal mechanism for `soft' hadron production in hadron-hadron, hadron-nucleus and nucleus-nucleus collisions. We emphasized that QCD does not work for these cases. So lacking a workable theory, we have relied on the phenomenology and all the salient features of our model fit very well with the experimental data. We hope that the work done here will throw enough light on the multiparticle production mechanism in nucleus-nucleus collisions. This also reveals that any deviations observed in the data from the predictions will help us in identifying QGP formation. Furthemore, we expect that this model can also be extended to accomodate the rapidity, pseudorapidity as well as transverse mass distribution of charged hadrons in the whole rapidity region which will be studied in a future publication.\\

\noindent
{\bf Acknowledgments}\\

 AK and PKS are grateful to the University Grants Commission (UGC), New Delhi for providing a research grant. BKS also sincerely acknowledges the Indian Space Research Organization (ISRO), India for providing the financial support.

\pagebreak

\end{document}